\documentclass{article}
\usepackage{amssymb}

\usepackage{ragged2e}       
\usepackage[protrusion=true,expansion=false]{microtype} 
\usepackage{graphicx}
\usepackage{multirow, multicol}
\usepackage{caption}
\usepackage{float}
\usepackage{booktabs, tablefootnote}
\usepackage[table,xcdraw]{xcolor}
\usepackage{amsthm, amsmath,amssymb,amsfonts}
\usepackage{mathtools} 
\usepackage{mathrsfs}
\usepackage{algorithm}
\usepackage{algpseudocode}  
\usepackage{algorithmicx}

\usepackage{textcomp}
\usepackage{pifont}
\usepackage{setspace,lineno}
\usepackage{listings}
\usepackage{soul}            
\sethlcolor{yellow}          
\usepackage[title]{appendix}
\usepackage[square,numbers]{natbib}
\usepackage{manyfoot}
\usepackage{xcolor}
\definecolor{customgreen}{HTML}{00B050}
\definecolor{captionblue}{HTML}{0070C0}  
\newcommand{\cmark}{\textcolor{green!60!black}{\ding{51}}} 
\newcommand{\xmark}{\textcolor{red}{\ding{55}}} 

\DeclareCaptionLabelFormat{boldblue}{\textbf{\textcolor{captionblue}{#1 #2}}}
\usepackage{titlesec}
\titleformat{\section}
  {\normalfont\Large\bfseries\color{captionblue}}{\thesection}{1em}{}
\titleformat{\subsection}
  {\normalfont\large\bfseries\color{captionblue}}{\thesubsection}{1em}{}
\titleformat{\subsubsection}
  {\normalfont\normalsize\bfseries\color{captionblue}}{\thesubsubsection}{1em}{}

\usepackage{geometry}
\usepackage{amsmath} 
\allowdisplaybreaks[4]
\usepackage[protrusion=true,expansion=false]{microtype}
\usepackage{hyperref}
\hypersetup{
    colorlinks=true,
    linkcolor=blue,
    citecolor=blue,
    urlcolor=blue
}
\title{ARIONet: An Advanced Self-supervised Contrastive Representation Network for Birdsong Classification and Future Frame Prediction}
\author{
Md. Abdur Rahman\textsuperscript{1}, 
Selvarajah Thuseethan\textsuperscript{2}, 
Kheng Cher Yeo\textsuperscript{2}, 
Reem E. Mohamed\textsuperscript{3},  
Sami Azam\textsuperscript{2,*}\\
\small
\textsuperscript{1}Department of Computer Science and Engineering, United International University, Dhaka, 1212, Bangladesh \\
\small
\textsuperscript{2}Faculty of Science and Technology, Charles Darwin University, Darwin, Northern Territory, 0909, Australia \\
\small
\textsuperscript{3}Faculty of Science and Information Technology, Charles Darwin University, Sydney, NSW, Australia \\
\small 
\textsuperscript{*} Corresponding Author: sami.azam@cdu.edu.au
}
\date{} 
\begin{document}
\justifying
\twocolumn[
\maketitle
\begin{abstract}
\noindent 
Automated birdsong classification is essential for advancing ecological monitoring and biodiversity studies. Despite recent progress, existing methods often depend heavily on labeled data, use limited feature representations, and overlook temporal dynamics essential for accurate species identification. In this work, we propose a self-supervised contrastive network, ARIONet (\underline{A}coustic \underline{R}epresentation for \underline{I}nterframe \underline{O}bjective \underline{Net}work), that jointly optimizes contrastive classification and future frame prediction using augmented audio representations. The model simultaneously integrates multiple complementary audio features within a transformer-based encoder model. Our framework is designed with two key objectives: (1) to learn discriminative species-specific representations for contrastive learning through maximizing similarity between augmented views of the same audio segment while pushing apart different samples, and (2) to model temporal dynamics by predicting future audio frames, both without requiring large-scale annotations. We validate our framework on four diverse birdsong datasets, including the British Birdsong Dataset, Bird Song Dataset, and two extended Xeno-Canto subsets (A-M and N-Z). Our method consistently outperforms existing baselines and achieves classification accuracies of 98.41\%, 93.07\%, 91.89\%, and 91.58\%, and F1-scores of 97.84\%, 94.10\%, 91.29\%, and 90.94\%, respectively. Furthermore, it demonstrates low mean absolute errors and high cosine similarity, up to 95\%, in future frame prediction tasks. Extensive experiments further confirm the effectiveness of our self-supervised learning strategy in capturing complex acoustic patterns and temporal dependencies, as well as its potential for real-world applicability in ecological conservation and monitoring.

\end{abstract}

\vspace{0.5em}
\noindent \textbf{Keywords:} Self-supervised; Contrastive Learning; Temporal Modeling; Future Frame; Birdsong; Acoustic Signal
\vspace{1em}
]

\section{Introduction}
\label{intro}
Birds are key ecological indicators whose presence, abundance, and vocal activity reflect the health of natural ecosystems. Birdsong plays a central role in avian communication, governing behaviors such as territorial defense, mating, and species recognition \cite{podos2022ecology}. However, many species of birds worldwide are currently in decline, with 12–13\% threatened with extinction due to habitat loss, climate change, and anthropogenic disturbance \cite{liu2022birdsongs, campbell2024birds}. Alarmingly, this decline affects not only rare species but also once-abundant birds on multiple continents \cite{reif2024accelerated, morrison2021bird}. For example, in Australia, the 2019-2020 mega fires alone severely impacted about 900 plant and animal species \cite{driscoll2024biodiversity}, contributing to more than 50\% of the national drop in Australia's avian red list index \cite{berryman2024trends}. 

As traditional field monitoring becomes impractical on a large scale, passive acoustic monitoring is increasingly used as a non-invasive and cost-effective method to track bird populations in real time \cite{ross2023passive}. However, these systems generate massive volumes of noisy, unstructured audio data, making automated birdsong classification a technical necessity and an ecological priority \cite{napier2024advancements}.

Due to the growing need for scalable biodiversity monitoring, researchers have developed various machine learning (ML) methods for birdsong classification. Early approaches relied on supervised learning with hand-crafted features such as Mel-frequency cepstral coefficients (MFCCs), chromagram, and spectral roll-off, paired with classical classifiers or Convolutional Neural Networks (CNNs) \cite{lakdari2024mel, michaud2023unsupervised}. Transfer learning later gained momentum, enabling models pretrained on large datasets to be fine-tuned for regional or low-resource settings. These approaches achieved strong performance in hundreds of bird species \cite{kahl2021birdnet}. More recent work introduced hybrid networks that fuse spectral and temporal cues, as well as compact architectures optimized for edge deployment \cite{gupta2021comparing, zhou2025mff}. Parallel to this, multi-feature fusion techniques combined mel-frequency cepstral coefficients, chromagram, and temporal statistics to improve noise robustness \cite{michaud2023unsupervised, ghani2025impact, quinn2022soundscape}. Chromagram-based and pitch-sensitive methods have also attracted attention for their ability to capture melodic structures. In parallel, self-supervised and Contrastive Learning frameworks have emerged, learning audio representations from unlabeled data through augmentation and sequence modeling \cite{wu2024orchard}. These models have demonstrated competitive accuracy across diverse habitats and species, enabling efficient and scalable monitoring without large annotation costs.

Despite these advances, several challenges remain in the current birdsong classification systems. Many supervised and transfer learning approaches are highly based on annotated data, which are costly and time-consuming to obtain, especially for rare or region-specific species \cite{ghani2025impact}. Models trained on spectrograms or static feature sets often struggle to capture the dynamic temporal structure of birdsong, particularly in real-world environments with overlapping calls, pitch shifts, and background noise \cite{hu2023deep, wang2024hierarchical}. Even in self-supervised settings, many recent methods prioritize global representations or single-view augmentations, which may overlook the fine-grained temporal and harmonic nuances crucial to species differentiation. Moreover, features like chromagram and mel-frequency cepstral coefficients are often treated as fixed inputs rather than evolving sequences, which might limit the model's ability to track pitch variation over time. These gaps make it difficult to develop systems that are accurate and robust in diverse acoustic conditions. 

To address these limitations, we propose a self-supervised framework for birdsong classification that learns temporally structured, pitch-sensitive representations from the chromagram-based audio input. Our method combines energy-based denoising, domain-specific augmentations, and a transformer-based backbone trained with both contrastive and predictive learning objectives. This design enables the model to learn fine-grained invariant features while preserving the sequential nature of bird singing. Unlike traditional spectrogram- or MFCC-based approaches, our framework treats chromagram features as dynamic sequences, which enhances sensitivity to pitch continuity and timing variations. Our model offers a scalable and robust solution for automated avian monitoring across diverse species and habitats by eliminating the need for extensive manual annotation and improving generalization to real-world acoustic conditions.

The major contributions of our study are as follows:
\begin{itemize}
    \item Proposed a dual-objective self-supervised learning framework that jointly optimizes contrastive learning and future-frame prediction. The model can learn both species-discriminative features and temporal dynamics directly from unlabeled data.
    \item A novel domain-specific augmentation strategy is introduced for birdsong classification, incorporating biologically grounded chromagram masking, pitch shifting, and time masking. This targeted scheme generates diverse acoustic views of the same signal and enhances the model’s ability to learn invariant representations under varying pitch, temporal distortions, and environmental noise conditions.
    \item A chromagram-centric representation is proposed to model birdsong as a temporal pitch-class sequence for future frame prediction. The design captures harmonic continuity and pitch stability and allows the model to focus on species-specific tonal patterns rather than broad spectral variations.
    \item A lightweight transformer-based encoder is developed to integrate multiple complementary audio features, such as mel-frequency cepstral coefficients, delta coefficients, chromagram short-time fourier transform (STFT), and spectral descriptors, into a unified and expressive sequence embedding.
    \item Comprehensive evaluations have been conducted on four diverse birdsong datasets in different audio formats to show the effectiveness of the proposed method compared to existing approaches. 
\end{itemize}

The rest of this paper is organized as follows. Section \ref{related_woks} reviews recent related studies on birdsong classification, self-supervised learning, and audio representation techniques. Section \ref{methodology} details the proposed methodology, including data set descriptions, audio pre-processing, feature extraction, and the design of the proposed framework. Section \ref{results_and_exp} presents the experimental results, including the performance of the model, ablation studies, and comparison with recent state-of-the-art methods. Section \ref{discussion} discusses the implications of the findings and potential future directions. Finally, Section \ref{conclusion} concludes the paper by summarizing the key contributions and outcomes.

\section{Related works} 
\label{related_woks}
In this section, we review recent work on automated birdsong classification. We cover traditional transfer learning methods, acoustic feature engineering, fusion techniques, and emerging self-supervised approaches that aim to capture temporal and harmonic structures in this domain.

\subsection{Traditional transfer learning approaches}
Early work in automatic birdsong recognition was mostly supervised and involved transfer learning. Kahl et al. \cite{kahl2021birdnet} introduced BirdNET, a CNN-based model built on a ResNet variant. It could identify more than 1000 bird species from spectrograms and reached a mean average precision of 0.791 on single species recordings. Transfer learning also showed great potential. Studies such as \cite{ghani2025impact, lauha2022domain, ghani2023global} tested different pretrained CNN backbones to see how well they generalize. For example, Ghani et al. \cite{ghani2023global} proposed global birdsong embeddings and found that models trained on large datasets perform much better than those trained from scratch, especially in low-resource settings. In another work, Ghani et al. \cite{ghani2025impact} used BirdNET knowledge distillation and reached an F1-score of 0.71. Gupta et al. \cite{gupta2021comparing} explored Recurrent CNNs for large-scale bird classification. Their hybrid networks performed better than traditional ImageNet-based models and scored 90\% accuracy in 100 bird species.

However, despite their successes, these approaches require extensive labeled data and often show reduced robustness in noisy or field-recorded environments. Additionally, their reliance on existing pretrained and transfer learning, as well as spectrogram-based CNNs, can limit sensitivity to fine-grained pitch information and may fail to capture the dynamic temporal structure of birdsong. 

\subsection{Feature extraction and multi-modal representations}
Another line of work focuses on engineering and the fusion of various acoustic features. Traditionally, many studies used classical descriptors such as mel-frequency cepstral coefficients, chromagram, and spectral statistics. For example, Lakdari et al. \cite{lakdari2024mel} showed that mel-frequency cepstral coefficients outperformed CNN-based embeddings in noisy conditions, especially for species-specific gibbon calls. Similarly, studies such as \cite{li2025multi, xie2022multi} emphasized multi-feature fusion, combining mel-frequency cepstral coefficients with chromagram and temporal stats to improve robustness. Likewise, Liu et al. \cite{liu2022birdsong} applied multi-feature channel fusion using 2D and 3D CNNs on log-mel-spectrograms and waveform images, achieving an mean average precision of 95.9\% across four orchard bird species. Although chroma-based and pitch-sensitive methods were less common, they are gaining attention. In particular, Ugarte et al. \cite{ugarte2024unveiling} highlighted the importance of chromagram mel-frequency cepstral coefficients and spectral roll-off, showing that pitch combinations improve generalization. Using 19 features in a heterogeneous subset, they achieved a precision of more than 95\% with a nearest-neighbor classifier. Meanwhile, Hu et al. \cite{hu2023deep} fused mel-frequency cepstral coefficients with an attention-based ResNet18 to better capture spectral and temporal cues. With early fusion, their MFF-ScSEnet reached 96.28\%–98.34\% accuracy across three datasets. Similarly, Wang et al. \cite{wang2023hierarchical} proposed a hierarchical model that combines static spectral and dynamic temporal features through sequential layers, achieving 93.67\%–97.02\% accuracy on the same datasets.

However, most of these studies rely on supervised training and hand-crafted feature fusion, with limited emphasis on learnable representations of pitch dynamics. They often treat chromagrams or mel-frequency cepstral coefficients as static features rather than modeling them as evolving temporal sequences. This simplification can interfere with the ability to capture the temporal complexities of birdsong.

\subsection{Self-supervised and contrastive learning}
Inspired by self-supervised contrastive learning frameworks such as SimCLR and wav2vec, many studies have adopted similar methods for birdsong and animal sound analysis. For example, DBS-NET \cite{wang2025dbs} combined supervised and self-supervised branches to learn dual representations. On both a custom 30-class dataset and the Birdsdata dataset, it reached an accuracy of 97.54\% and 97.09\%, respectively. Meanwhile, cross-domain studies also highlighted generalization via self-supervision. For example, Michaud et al. \cite{michaud2023unsupervised} used unsupervised clustering to refine noisy labels, while Zhong et al. \cite{zhong2020multispecies} applied pseudo-labeling in a transfer learning setup, reaching 97.7\% sensitivity and 96.4\% specificity for 24 species. Finally, Wu et al. \cite{wu2024orchard} applied multi-level contrastive learning for orchard bird recognition, fusing temporal and frequency features, and achieved 99.40\% and 92.67\% accuracy on the Orchard-birds and Birdsdata datasets, respectively.

Although promising, many of these methods either focus solely on global representations or treat time-frequency features as static inputs. In addition, they often overlook the temporal continuity and dynamic nature of birdsong, which can limit their ability to model sequential vocal patterns. In addition, few approaches combine contrastive learning with sequence-based prediction tasks. These methodological gaps suggest the need for frameworks that jointly capture both invariant representations and temporal dependencies in birdsong data. Thus, to address these aforementioned issues, we proposed a self-supervised learning framework that unifies contrastive representation learning with future-frame prediction to capture both invariant species-specific features and the temporal dynamics of birdsong. 

\section{Methodology} \label{methodology}
\begin{figure*}[h]
\centering
\includegraphics[scale=0.04]{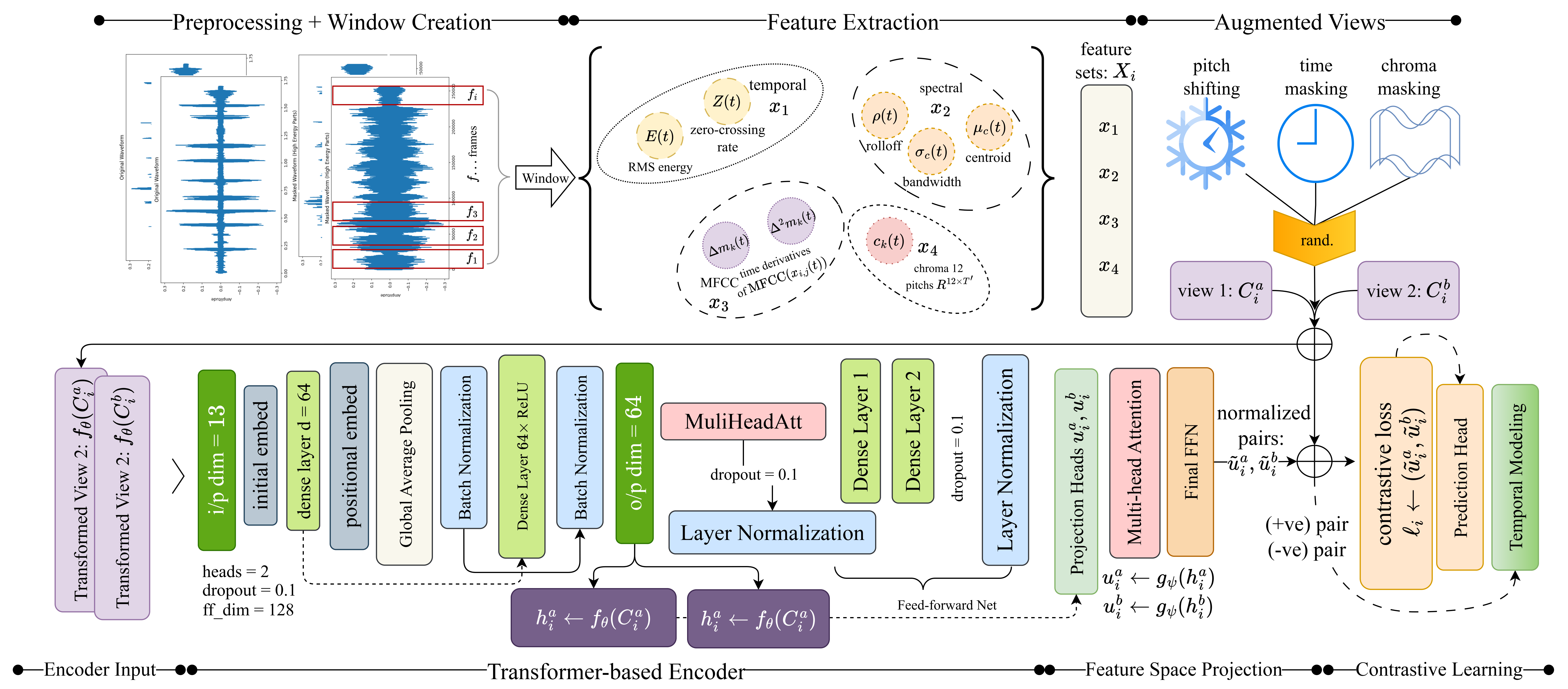} 
\caption{Overview of the proposed framework. Processed samples are segmented and converted into 4 feature types: temporal, spectral, MFCC, and chromagram. Augmented views are created using pitch shifting, time masking, and chromagram masking, then encoded via a shared transformer with positional embeddings and multi-head attention. Then the projected embeddings are optimized using contrastive and temporal prediction losses.}
\label{fig:ARIONet_main}
\end{figure*}

The goal of this study is to develop a self-supervised framework that learns discriminative species-specific patterns and temporal dynamics representations from birdsong audio data to support two downstream tasks: (1) species classification and (2) future frame prediction. Subsequent sections detail the pipeline, and Figure \ref{fig:ARIONet_main} summarizes the proposed pipeline.

\begin{table*}[h]
\centering
\caption{Summary of the four birdsong audio datasets used in this study. To ensure consistency in terms of window size (for each dataset), we calculated the minimum number of windows per species (mWin/Sp), and this number was used to create the total windows for experiments. For the Xeno-Canto subsets, the number of windows varies due to the organized structure. For better readability, each dataset is assigned a short reference name (see the \textbf{Referred} column) used throughout the paper.}
\label{tab:dataset_info}
\renewcommand{\arraystretch}{1.2}
\small
\begin{tabular}{cllccccc}
\midrule
\textbf{Ref.} & \textbf{Datasets} & \textbf{Referred} & \textbf{Specie} & \textbf{Sample} & \textbf{Format} & \textbf{mWin/Sp} & \textbf{Windows} \\
\midrule
\cite{d1british_birdsong} & British Birdsong Dataset & XC-British                 & 85 & 264 & .flac &  20   & 18386 \\
\cite{d2shanbhag_bird_song} & Bird Song Dataset & XC-BS5 & 5 & 5422  & .wav       &  3498 & 21772 \\
\cite{d3rao_a2m} & Xeno-Canto Bird Recordings Extended A-M & XC A-M & 153 & 14685 & .mp3 &  Varies   & Varies \\
\cite{d4rao_n2z} & Xeno-Canto Bird Recordings Extended N-Z & XC N-Z & 106 & 9099 & .mp3 &  Varies   & Varies \\
\midrule
\end{tabular}
\vspace{-15pt}
\end{table*}

\vspace{-15pt}
\subsection{Datasets} 
\label{dataset}
In this study, we used four publicly available birdsong audio datasets, originating from the Xeno-Canto\footnote{https://xeno-canto.org/} collection, that vary in terms of species diversity, a broad spectrum of birdsong characteristics, annotation quality, and recording conditions. Table \ref{tab:dataset_info} summarizes key statistics for each data set, including the number of species, audio format, sample counts, and the number of fixed time windows derived per species.

The British Birdsong Dataset \cite{d1british_birdsong} includes high-quality Free Lossless Audio Codec (FLAC) recordings from 85\footnote{Total number of unique species as per the source was 88; however, only 85 species had associated labels.} labeled species, with 264 labeled audio samples segmented into 18,386 fixed-length windows, using a cap of 20 windows per species to reduce class imbalance. The Bird Song Dataset \cite{d2shanbhag_bird_song} comprises Waveform Audio File Format (WAV) recordings from five species, with 5,422\footnote{The dataset was originally sourced for 9107 samples, but 5422 were labeled} labeled samples contributing 21,772 windows, up to 3,498 per species. The extended Xeno-Canto Bird Recordings dataset is organized into two main subsets: one for species from A to M \cite{d3rao_a2m} and another for species from N to Z \cite{d4rao_n2z}. The subsets span 153 and 106 species, respectively, and contain MP3 recordings of varying duration and quality. Within each of these subsets, there are subdirectories named after the scientific names of the bird species. These two datasets required extensive preprocessing, with the number of extracted windows varying significantly due to inconsistent recording lengths and species distribution due to their size. The subsets \cite{d3rao_a2m, d4rao_n2z} contain 23,784 valid files, organized alphabetically into subdirectories based on the initial letters of the species names (see Section \ref{traning_analysis} for details).

It is worth noting that in \cite{d4rao_n2z}, species with \textit{q}, \textit{u}, \textit{x}, and \textit{z} initials had no samples; thus, we continued with the rest of the directories. For simplicity, we refer to the datasets as XC-British, XC-BS5, XC A-M, and XC N-Z, respectively, in the following sections (see Table \ref{tab:dataset_info}). 

\subsection{Problem formulation}
Let $x(t)$ be a raw birdsong waveform of arbitrary duration $T$, drawn from a labeled dataset $\mathcal{D}_{\text{labeled}}$, which contains $x(t)$ and the species identity $y(t)$. Each waveform $x(t)$ is segmented into overlapping fixed-length frames using a sliding window to enable learning of a structured representation. For each frame, we extract a comprehensive multiview acoustic representation $[\mathbf{x}_1, \dots, \mathbf{x}_i]$, where each frame-level vector $\mathbf{x}_i$ includes mel-frequency cepstral coefficients, delta and delta-delta mel-frequency cepstral coefficients, chromagram Short-Time Fourier Transform, and spectral descriptors such as centroid, bandwidth, roll-off, Root mean square (RMS) energy, and zero-crossing rate. In addition, a chromagram tensor $\mathbf{C}$ is calculated per time window to support pitch-class modeling over local sequences.

We formulate two complementary pretext tasks: a contrastive task and a predictive task. For the former, we generate two augmented views (i.e., $x_i^a, x_i^b$) of the same audio segment using domain-specific perturbations such as pitch shifting, time masking, and chromagram masking. A transformer-based encoder is used $f_{\theta}$ to map each view to a latent representation, and a contrastive loss of $f_{\theta}(x_i^a), f_{\theta}(x_i^b)$ is used to maximize the agreement between positive pairs while separating negatives in the latent space. To model temporal dynamics, we introduce a predictive task that captures the evolution of frame-level features. Given a sequence of previous feature vectors $[\mathbf{f}_{t-\tau}, \dots, \mathbf{f}_{t}]$ in $F$, the model learns to predict a future feature vector $\mathbf{f}_{t+\delta}$ and minimize the error loss. This encourages the encoder to learn temporally coherent representations.

\subsection{Preprocessing}
\label{preproc}
To remove silent or low-energy portions that are unlikely to contain useful birdsong, we applied a simple but effective energy-based filtering step. Starting with the raw waveform $x(t)$, we compute its Mel-spectrogram $S$, where each column represents the energy distribution across the mel frequency bins for a short time frame. For each frame $n$, we calculate the mean spectral energy, as shown in Equation \eqref{eq: mean_spectral_energy}:
\begin{equation}
    \bar{S}_n = \frac{1}{F} \sum_{f=1}^{F} S_{f,n}
    \label{eq: mean_spectral_energy}
\end{equation}

\begin{figure}[h]
\centering
\includegraphics[scale=0.06]{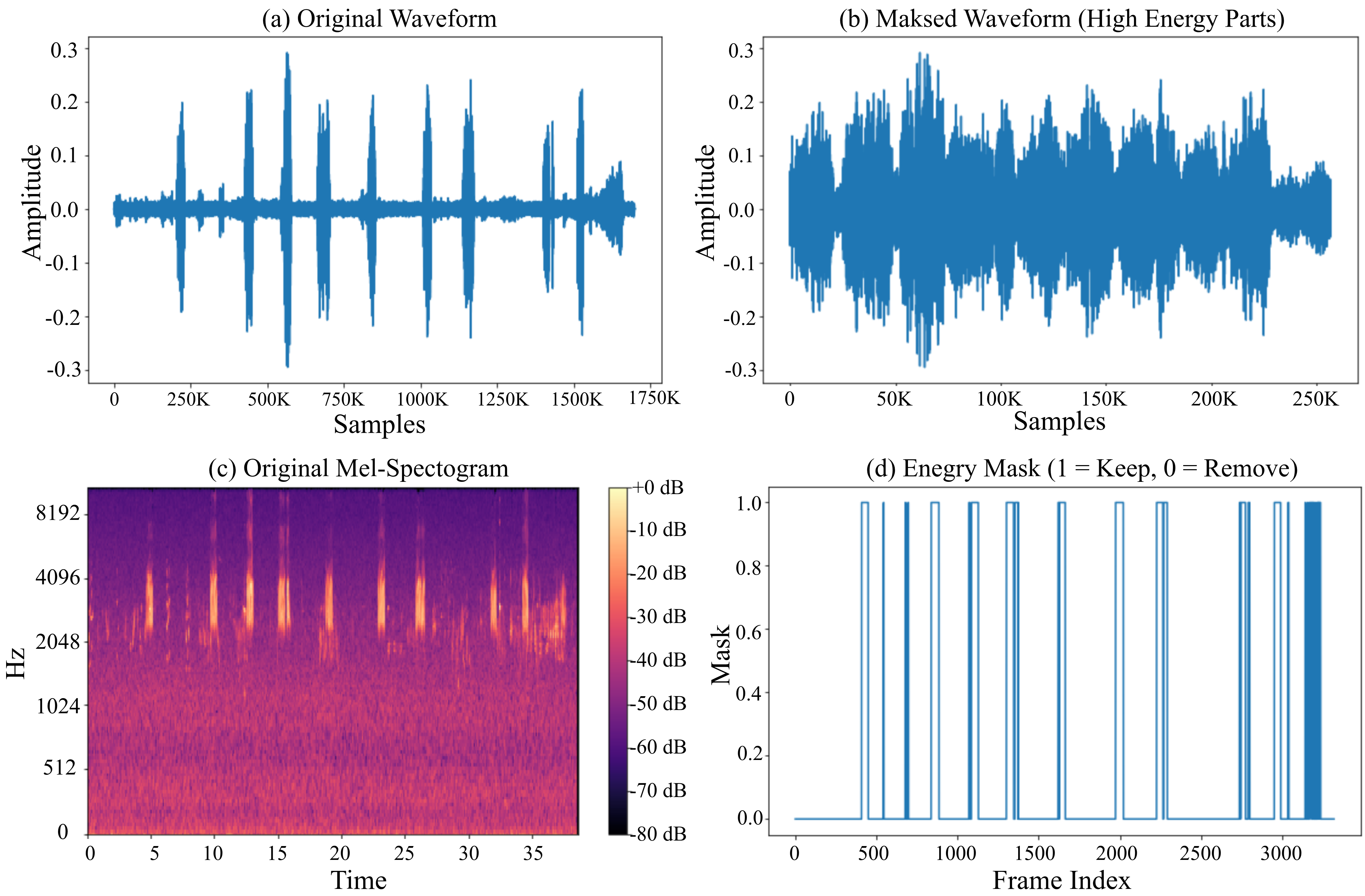}
\caption{Visualization of an original audio sample along with the results of 5\% low-energy filtering. (a) shows the original waveform. (c) displays the original Mel-Spectrogram. The resulting (d) Energy Mask identifies frames to keep (1) or remove (0) based on the 5\% threshold of the peak frame energy. The resulting masked waveform, containing only the high-energy segments, is shown in (b).}
\label{fig:energy_filter}
\end{figure}

We then identify the frame $n^*$ with the highest average energy $M$ and use its value, $\bar{S}_{n^*}$, as a reference. Any frame with energy below $M / 20$, that is, less than 5\% of the peak frame energy, is considered low energy and discarded (see Figure \ref{fig:energy_filter}). This frame-level mask is projected back to the waveform using the spectrogram's hop length to allow us to construct a sample-level mask that keeps only the most active segments. To ensure consistency across samples with varying durations, we dynamically set the window size, the length of a contiguous segment of the waveform, to the minimum audio length observed in the dataset. Each waveform is split into non-overlapping windows of equal size, which enables localized time-frequency analysis across different parts of the recording and guarantees that every sample contains at least one valid window. 
It should be noted that some segments are discarded if they do not meet the minimum criteria for feature computation. Specifically, we skip any windowed segment that cannot produce at least 13 chromagram frames. Each chromagram frame represents energy across 12 pitch classes\footnote{The 12 musical pitch classes are: C,\allowbreak{} C$\sharp$\allowbreak{}/D$\flat$\allowbreak{},\allowbreak{} D,\allowbreak{} D$\sharp$\allowbreak{}/E$\flat$\allowbreak{},\allowbreak{} E,\allowbreak{} F,\allowbreak{} F$\sharp$\allowbreak{}/G$\flat$\allowbreak{},\allowbreak{} G,\allowbreak{} G$\sharp$\allowbreak{}/A$\flat$\allowbreak{},\allowbreak{} A,\allowbreak{} A$\sharp$\allowbreak{}/B$\flat$\allowbreak{}, and \allowbreak{} B.}, capturing the harmonic content of the signal. A sequence of 13\footnote{Here, the 13-frame requirement comes from the temporal axis. The chromagram features form a shape matrix (12, T), where 12 denotes the pitch classes and T is the number of time frames.} such frames (a chromagram matrix \(12\times13\)) ensure a brief but musically meaningful span. Species with no valid windows remaining after this filtering are excluded from training. 

\subsection{Feature extraction}
Following preprocessing, each windowed audio segment \( x \) is transformed into a structured time-frequency representation through a series of audio features that capture the signal’s spectral, timbral, and harmonic characteristics. These features are computed frame-wise and aggregated to form a consistent matrix of \(\phi(x)^{F \times T} \), where \( F \) denotes the number of feature channels and \( T \) represents the number of time frames retained after truncation. Algorithm \ref{alg:algo_preprocessing} outlines the preprocessing and feature extraction process.

\begin{algorithm}[ht]
\caption{Birdsong preprocessing and feature extraction pipeline.}
\label{alg:algo_preprocessing}
\textbf{Input:} raw waveform dataset $\mathcal{D}$ of $x_i(t), y_i(t)$,  
window length $L$, hop size $H$,  
energy threshold ratio $\tau$, minimum chromagram length $T_{\min}$

\begin{algorithmic}[1]
\For{each waveform $x_i(t) \in \mathcal{D}$}

    \State \textit{\color{gray}// preprocessing}
    \State mel-spectrogram: $S_i \gets \text{MelSpec}(x_i(t))$
    \State frame energy: $e_n \gets \text{mean}(S_i[:, n])$ for all frames $n$
    \State \textit{\color{gray}// extracting high-energy regions} 
    \State threshold: $\epsilon \gets \tau \cdot \max_n e_n$
    \State high-energy frames: $\mathcal{F}_i \gets \{n \mid e_n \geq \epsilon\}$
    \State extract non-silent region $x_i^{\text{eff}}(t)$ corresponding to $\mathcal{F}_i$
    \State segment $x_i^{\text{eff}}(t)$ into $L$ with $H$

    \For{each segment $x_{i,j}(t)$}
        \State compute chromagram: $C_{i,j} \gets \text{chroma}(x_{i,j}(t))$
        \State \textit{\color{gray}// chromagram segments filtering} 
        \If{$\text{len}(C_{i,j}) < T_{\min}$}
            \State \textbf{continue} 
        \EndIf
        
        \State \textit{\color{gray}// extracting features} 
        \State compute MFCCs: $M \gets \text{MFCC}(x_{i,j}(t))$
        \State compute deltas: $\Delta M, \Delta^2 M$
        \State spectral features: centroid, bandwidth, roll-off
        \State temporal features: RMS energy, zero-crossing rate
        \State concat features to form $z_{i,j}$
        \State append $z_{i,j}$ to $\mathcal{Z}$ and $C_{i,j}$ to $\mathcal{C}$
    \EndFor
\EndFor

\State \Return $\mathcal{Z}, \mathcal{C}$
\end{algorithmic}
\textbf{Output:} feature matrix set $Z_i$, chromagram set $C_i$
\end{algorithm}

\vspace{0.02\linewidth}\noindent\textbf{Mel-frequency cepstral coefficients (MFCCs).} 
We begin by computing the mel-frequency cepstral coefficients, which characterize the short-term spectral envelope of the signal by projecting the log-mel-spectrogram into a correlated space. For each frame \( t \), we extract 13 base mel-frequency cepstral coefficients, denoted as \( m_k(t) \), where \( k = 1, 2, \dots, 13 \). The MFCC matrix is thus calculated as shown in Equation \eqref{eq:mfcc}:
\begin{equation}
    \text{MFCC}(x) = m_1(t), m_2(t), \dots m_{13}(t)]^\top \in \mathbb{R}^{13 \times T}
    \label{eq:mfcc}
\end{equation}
where $m_k(t)$ is the $k^{th}$ mel-frequency cepstral coefficients coefficient in time frame $t$, and $T$ is the number of frames in the segment. Then, to capture local temporal dynamics, we compute the first- and second-order time derivatives of each mel-frequency cepstral coefficients using finite differences following Equation \eqref{eq:delta_mfcc}: 
\begin{equation}
\begin{aligned}
    \Delta m_k(t) &= m_k(t) - m_k(t - 1) \\
    \Delta^2 m_k(t) &= \Delta m_k(t) - \Delta m_k(t - 1)
\end{aligned}
\label{eq:delta_mfcc}
\end{equation}

where $\Delta m_k(t)$ and $\Delta^2 m_k(t)$ denote the velocity and acceleration of coefficient $m_k$ at frame $t$, respectively. When concatenated with the original mel-frequency cepstral coefficients, these yield a 39-dimensional descriptor per frame, which is then averaged over time to produce a fixed-length feature vector for the entire segment.

\vspace{0.02\linewidth}\noindent\textbf{Spectral features.}
In addition to mel-frequency cepstral coefficients, we extract a suite of spectral features designed to characterize the energy distribution and shape of the signal's power spectrum. Let $S_{f,t}$ denote the magnitude of the spectrogram in the frequency bin $f$ and the time frame $t$. The spectral centroid $\mu_c(t)$, the bandwidth $\sigma_c(t)$, and the roll-off frequency $\rho(t)$ are defined as follows:
\begin{equation}
\mu_c(t) = \frac{\sum_{f} f \cdot S_{f,t}}{\sum_{f} S_{f,t}}
\label{eq:spectral_centroid}
\end{equation}
\begin{equation}
\sigma_c(t) = \sqrt{\frac{\sum_{f} (f - \mu_c(t))^2 \cdot S_{f,t}}{\sum_{f} S_{f,t}}}
\label{eq:spectral_bandwidth}
\end{equation}
\begin{equation}
    \rho(t) = \min \left\{ f : \sum_{f'=0}^{f} S_{f',t} \geq 0.85 \sum_{f'} S_{f',t} \right\}
\label{eq:spectral_rolloff}
\end{equation}
In Equations \eqref{eq:spectral_centroid}-\eqref{eq:spectral_rolloff}, $\mu_c(t)$ measures the center of mass of the spectrum, $\sigma_c(t)$ quantifies its spread, and $\rho(t)$ gives the frequency below which 85\% of the total energy is concentrated. These are computed for each frame and averaged to obtain a global summary of the segment.

\vspace{0.02\linewidth}\noindent\textbf{Root mean square energy and zero-crossing rate.} 
In parallel, the temporal energy and waveform periodicity are captured through the root mean square energy and zero-crossing rate. For a frame of $N$ samples, we compute the short-term energy, $E(t)$, of the frame and the rate at which the waveform crosses zero amplitude, $Z(t)$, following Equations \eqref{eq:rms}, and \eqref{eq:zcr}:
\begin{equation}
    E(t) = \sqrt{\frac{1}{N} \sum_{n=1}^{N} x_n^2}
\label{eq:rms}
\end{equation}
\begin{equation}
    Z(t) = \frac{1}{N-1} \sum_{n=1}^{N-1} i \left[ x_n x_{n+1} < 0 \right]
\label{eq:zcr}
\end{equation}
where $i[\cdot]$ is the indicator function. These features are also averaged across frames to form segment-level descriptors.

\vspace{0.02\linewidth}\noindent\textbf{Chromagram.} 
Finally, to capture harmonic content and pitch salience, we compute chromagram features by projecting the spectral energy onto 12 pitch classes corresponding to the semitones of the chromatic scale. For each time frame $t$, the chromagram vector $c_t \in \mathbb{R}^{12}$ is given in Equation \eqref{eq:chroma}:
\begin{equation}
c_{k}(t) = \sum_{f \in \mathcal{F}*k} S_{f,t} \quad \text{for } k = 1, 2, \dots, 12
\label{eq:chroma}
\end{equation}
where $S_{f,t}$ is the magnitude of the spectrogram in the frequency bin $f$, and $\mathcal{F}_k$ denotes the set of bins assigned to pitch class $k$. The resulting chromagram matrix $C$ is shown in Equation \eqref{eq:chroma_matrix}:
\begin{equation}
C = [c_1, c_2, \dots, c_{T'}] \in \mathbb{R}^{12 \times T'}
\label{eq:chroma_matrix}
\end{equation}
where $T'$ is the number of frames retained. Each column $c_t$ captures the normalized pitch energy in the frame $t$.

\begin{figure}[h]
\centering
\includegraphics[scale=0.048]{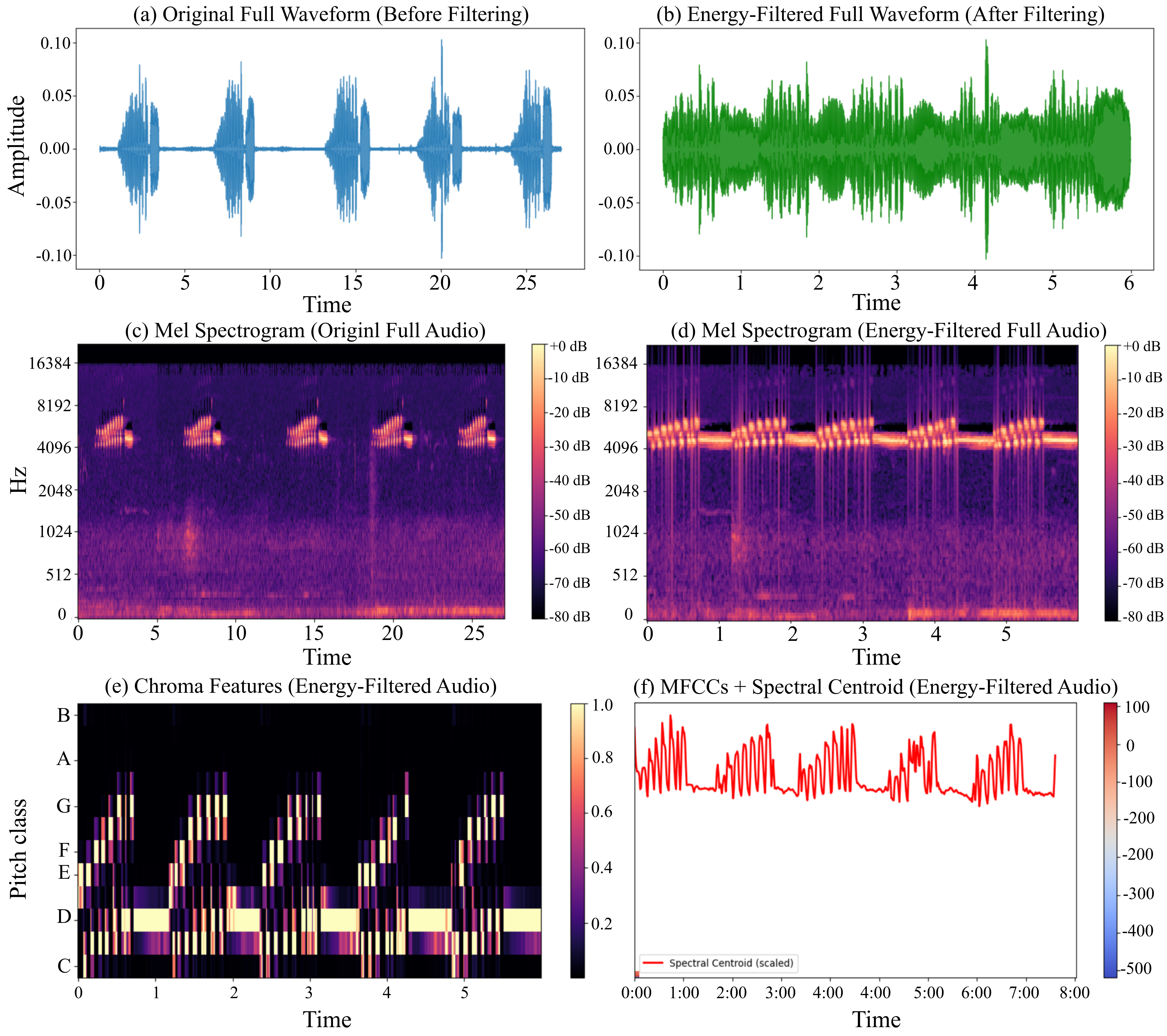} 
\caption{The sequence of processing and feature extraction is shown by: (a) the original full waveform (before filtering), which is transformed into (b) the energy-filtered full waveform. The corresponding time-frequency visualizations are (c) the Mel Spectrogram (original audio) and (d) the Mel Spectrogram (energy-filtered signal). Finally, the extracted features from the filtered signal include harmonic content with chroma features (e), and (f) mel-frequency cepstral coefficients and the spectral centroid (energy-filtered signal).}
\label{fig:features}
\end{figure}

All parameters, such as frame length and filter bank resolution, are dynamically adapted based on the sampling rate and effective window duration. The resulting representation $\phi(x)^{F \times T}$ serves as the unified input to our self-supervised learning framework, where temporal and frequency-based signals contribute to downstream discriminability. Figure \ref{fig:features} visualizes the features and processed output.

\subsection{Contrastive learning with multiview chromagrams}
To learn robust, structure-aware representations of birdsong in a self-supervised manner, we employ a contrastive learning framework grounded in multiview similarity over chromagram-based descriptors. The central idea is to leverage carefully constructed positive pairs derived from domain-specific augmentations of the same audio segment, encouraging the model to discover invariant patterns that persist across time-frequency transformations. 

For a collection of fixed-length birdsong segments, we map each segment $x_i$ through a feature extractor $\phi(\cdot)$ to a chromagram representation $\phi(x_i)^{F \times T}$, where $F$ denotes the number of chromagram bands and $T$ is the temporal length. These chromagrams serve as the base input for contrastive pretraining. The process is outlined in Algorithm \ref{alg:algo_contrastive}.

\begin{algorithm}[ht]
\caption{Contrastive learning on multiview chromagram features.}
\label{alg:algo_contrastive}
\textbf{Input:} Chromagram set $\mathcal{C}$,  
Augmentation functions $\mathcal{A}$,  
Transformer encoder $f_\theta$, projection head $g_\psi$,  
Temperature $\tau$, batch size $B$

\begin{algorithmic}[1]
\State \textit{\color{gray}// processing chromagram multiview} 
\For{each chromagram $C_i \in \mathcal{C}$}
    \State generate aug views: $C_i^a, C_i^b \gets \mathcal{A}(C_i)$
    \State encode w/ transformer: $h_i^a \gets f_\theta(C_i^a)$, $h_i^b \gets f_\theta(C_i^b)$
    \State project to feature space: $u_i^a \gets g_\psi(h_i^a)$, $u_i^b \gets g_\psi(h_i^b)$
    \State normalize: $\tilde{u}_i^a \gets \text{norm}(u_i^a)$, $\tilde{u}_i^b \gets \text{norm}(u_i^b)$
    \State store $(\tilde{u}_i^a, \tilde{u}_i^b)$ in batch
\EndFor
\State \textit{\color{gray}// initializing loss} 
\State loss $\mathcal{L}_{\text{con}} \gets 0$
\For{each pair $(\tilde{u}_i^a, \tilde{u}_i^b)$ in batch}
    \State \textit{\color{gray}// calculating similarity-dissimilarity aug} 
    \State identify positive and negative pairs
    \State compute similarity scores across batches
    \State compute contrastive loss $\ell_i$ for $(\tilde{u}_i^a, \tilde{u}_i^b)$
    \State update: $\mathcal{L}_{\text{con}} \gets \mathcal{L}_{\text{con}} + \ell_i$
\EndFor

\State avg loss: $\mathcal{L}_{\text{con}} \gets \mathcal{L}_{\text{con}} / B$
\State backpropagate and update $\theta, \psi$
\State \Return trained encoder $f_\theta$ and projection head $g_\psi$
\end{algorithmic}
\textbf{Output:} Learned representations via $f_\theta$, optimized for contrastive alignment
\end{algorithm}

\vspace{0.02\linewidth}\noindent\textbf{Multiview construction and transformer-based encoding.}
For each sample $c_i$, we generate two stochastic views $(c_i^a, c_i^b)$ by applying independent domain-specific augmentations: chromagram masking, time masking, and pitch shifting. These transformations preserve the semantic identity of the vocalization while perturbing its surface appearance, forming the basis for our view-level invariance assumption.

Each view is processed by an encoder $f_\theta$, which is implemented as a lightweight transformer stack. Specifically, $f_\theta: \mathbb{R}^{F \times T}$ maps the chromagram sequence to a contextualized embedding sequence via stacked self-attention blocks and feed-forward layers. The transformer architecture enables the model to attend to global temporal dependencies across the chromagram timeline, which is especially beneficial for capturing periodic and harmonic motifs characteristic of birdsong. We apply temporal average pooling to retain a smoothed temporal signature across the sequence and to derive a compact vector representation for each view. This also ensures that all time steps contribute equally. This is formalized in Equation \eqref{eq: temporal_avg_pool}:
\begin{equation}
    \begin{aligned}
        h_i^a = \text{AvgPool}(f_\theta(c_i^a)) \in \mathbb{R}^d, \\ 
        h_i^b = \text{AvgPool}(f_\theta(c_i^b)) \in \mathbb{R}^d
    \end{aligned}
\label{eq: temporal_avg_pool}
\end{equation}

These representations are then passed through a projection head $g_\psi: \mathbb{R}^d \rightarrow \mathbb{R}^{d'}$, producing the final embeddings for contrastive comparison, as shown in Equation \eqref{eq: final_emb}:
\begin{equation}
    u_i^a = g_\psi(h_i^a), \quad u_i^b = g_\psi(h_i^b)
\label{eq: final_emb}
\end{equation}
We then normalize each $u$ to a unit length $\tilde{u}$. These augmentations preserve the global temporal structure while introducing localized distortions, enabling the encoder to learn consistent long-range dependencies across time.

\begin{figure}[h]
\centering
\includegraphics[scale=0.075]{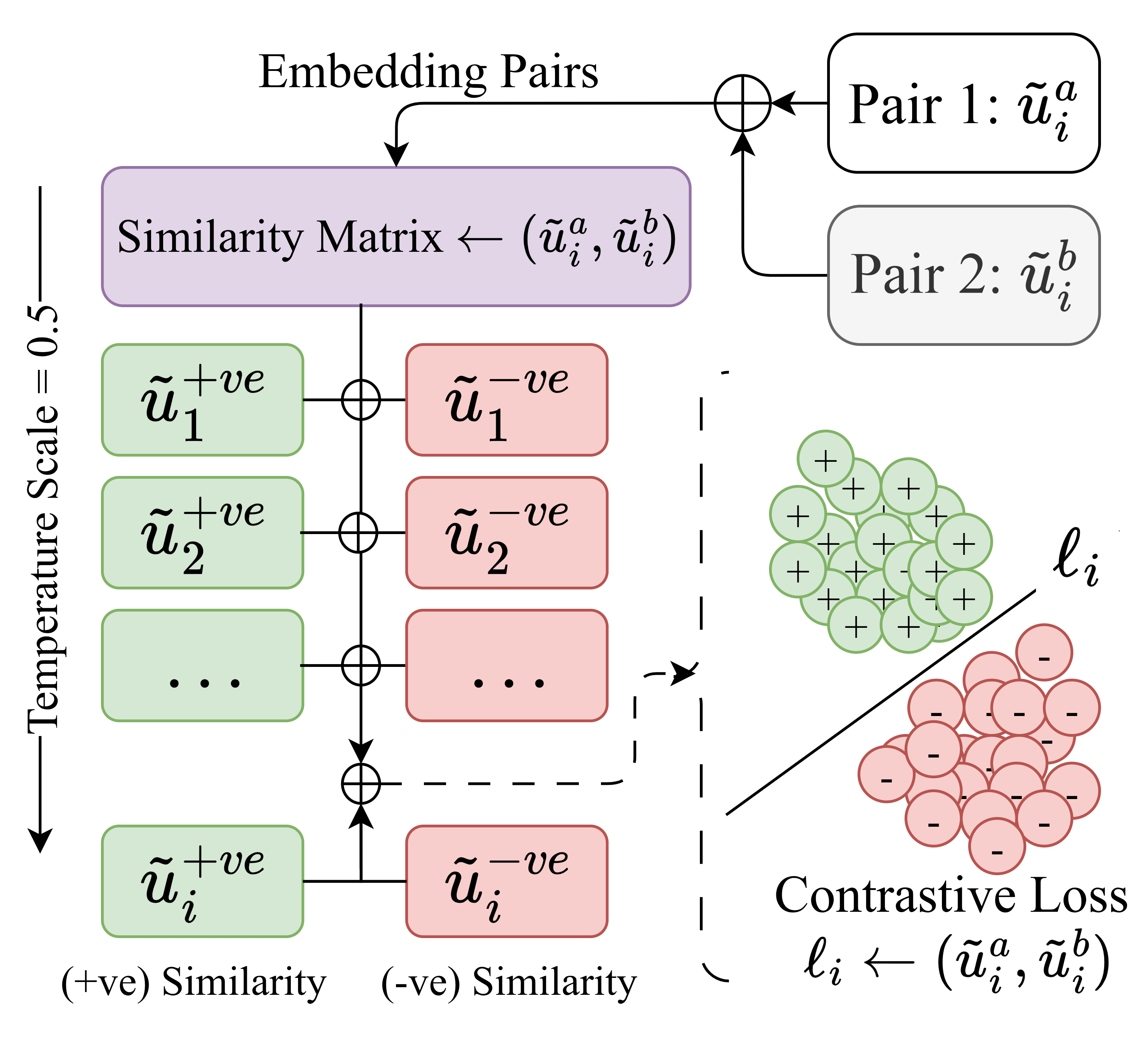} 
\caption{Contrastive learning module: embeddings are projected and normalized, then compared using cosine similarity to form a similarity matrix. The temperature-scaled loss, $\ell'_i$, pulls positive pairs together and pushes negative pairs apart for discriminative representations.}
\label{fig:contrastive_process}
\vspace{-15pt}
\end{figure}

\vspace{0.02\linewidth}\noindent\textbf{Contrastive similarity and loss formulation.}
The objective is to maximize the similarity between two views of the same sample while contrasting them against all other views in the batch. Let $u^\top v$ denote the cosine similarity. For a batch of $B$ samples (yielding $2B$ views), we define the positive pair for index $i$ as $(u_i^a, u_i^b)$, and treat all other views as negatives. The contrastive loss for each anchor $u_i^a$ with its positive $u_i^b$ is calculated as shown in Equation \eqref{eq:cl_loss_clean}:
\begin{equation}
\ell_i = -\log
\frac{\exp\!\big(\mathrm{sim}(\tilde{u}_i^a,\tilde{u}_i^b)/\tau\big)}
{\displaystyle
  \sum_{j=1}^{B}\exp\!\big(\mathrm{sim}(\tilde{u}_i^a,\tilde{u}_j^b)/\tau\big)
  +\sum_{\substack{j=1\\ j\ne i}}^{B}\exp\!\big(\mathrm{sim}(\tilde{u}_i^a,\tilde{u}_j^a)/\tau\big)
}
\label{eq:cl_loss_clean}
\end{equation}
where $\tau > 0$ is a temperature parameter that sharpens similarity scores and $i[\cdot]$ is an indicator function. The total loss is then symmetrized between both views following Equation \eqref{eq: total_loss}:
\begin{equation}
    \mathcal{L}_{\text{con}} = \frac{1}{2B} \sum_{i=1}^{B} \left( \ell_i + \ell'_i \right)
\label{eq: total_loss}
\end{equation}
where $\ell'_i$ corresponds to the reverse pair using $u_i^b$ as anchor and $u_i^a$ as its positive. The process is illustrated in Figure \ref{fig:contrastive_process}.

\vspace{0.02\linewidth}\noindent\textbf{Chroma-aware temporal representation learning.}
Through this formulation, the transformer encoder learns to produce embeddings that are invariant to chromagram-level perturbations while remaining sensitive to the temporal-harmonic structure of the underlying birdsong. The attention mechanism enables it to align semantically related spectral events across time, even when localized distortions (e.g., pitch transposition, amplitude envelope variations, and missing harmonic content) are present in the views.

In contrast to standard contrastive learning pipelines that operate on raw waveforms or mel-frequency cepstral coefficients, our formulation exploits the pitch-class aligned structure of chromagrams and models cross-time interactions using transformers. The resulting embedding space reflects meaningful vocal characteristics, such as motif repetition, harmonic texture, and melodic arc, without requiring any labels, thus laying a strong foundation for downstream classification, clustering, or sequence modeling tasks.

\subsection{Predictive temporal modeling via future frame prediction}
For the future frame prediction task, we incorporate a predictive objective that trains the model to anticipate future chromagram frames given a past context window. 

We work with feature sequences $z_i \in \mathbb{R}^{F \times T}$, where $F$ is the number of pitch-class bands and $T$ is the temporal length. For each sample, we split the sequence into a context segment and a prediction target. Given a context window of length $t$ and a prediction horizon of $k$ frames, we define the input context $z_i^{\text{ctx}}$ as the first $t$ columns of $z_i$, specifically $z_i[:,:t]$. The future target segment $z_i^{\text{fut}}$ is taken from the next $k$ frames, corresponding to $z_i[:,t:t+k]$. Next, we pass the context segment $z_i^{\text{ctx}}$ through a shared transformer encoder $f_\theta$, which produces a contextual representation $h_i^{\text{ctx}} \in \mathbb{R}^{d \times t'}$, where $t' \leq t$ reflects potential downsampling due to attention pooling. We then decode this representation using a lightweight prediction head $d_\phi$, which maps from $\mathbb{R}^{d \times t'}$ to $\mathbb{R}^{F \times k}$. This produces the predicted chromagram sequence $\hat{z}_i^{\text{fut}}$, computed by applying $d_\phi$ to $h_i^{\text{ctx}}$. The future prediction objective is formulated as an Mean Squared Error loss between the predicted frames and the actual future frames, defined in Equation \eqref{eq: future_pred_obj}:
\begin{equation}
    \mathcal{L}_{\text{pred}} = \frac{1}{N} \sum_{i=1}^{N} \left\| \hat{z}_i^{\text{fut}} - z_i^{\text{fut}} \right\|_2^2.
\label{eq: future_pred_obj}
\end{equation}

Following this, we combine the contrastive loss $\mathcal{L}_{\text{con}}$ and the predictive loss $\mathcal{L}_{\text{pred}}$ into a unified training objective to jointly optimize both representational invariance and temporal structure. The encoder is encouraged to model both local harmonic continuity and global structural transitions by predicting the evolution of pitch-class patterns.

\section{Results and experiments} 
\label{results_and_exp}

\subsection{Evaluation metrics}
We used a combination of standard classification metrics  and regression-based similarity measures to assess both classification and future-frame prediction tasks. For the classification task, we evaluate performance using accuracy, precision, recall (i.e., sensitivity), F1-score, specificity, mean absolute error (MAE), negative predictive value (NPV), false positive rate (FPR), and false negative rate (FNR). The model is further assessed using Cohen's Kappa and Matthews Correlation Coefficient (MCC), which provide balanced measures even under class imbalance (see Equations \eqref{eq:mcc} and \eqref{eq:kappa}). Here, $TP$, $TN$, $FP$, and $FN$ denote the true positives, true negatives, false positives, and false negatives, respectively.
\begin{equation}
\text{MCC} = \frac{TP \cdot TN - FP \cdot FN}{\sqrt{(TP + FP)(TP + FN)(TN + FP)(TN + FN)}}
\label{eq:mcc}
\end{equation}
\begin{equation}
\kappa = \frac {p_o - p_e}{1 - p_e}
\label{eq:kappa}
\end{equation}

In Equation \eqref{eq:kappa}, $p_o$ is the observed agreement between the predicted and true labels, and $p_e$ is the expected agreement by random chance. For the future frame prediction task, we report the cosine similarity between the predicted and ground-truth feature vectors, which evaluates the directional alignment of the high-dimensional spectral representation and is defined in Equation \eqref{eq:cosine_sim}:
\begin{equation}
\text{Cosine Similarity} = \frac{\mathbf{z}_t \cdot \hat{\mathbf{z}}_t}{| \mathbf{z}_t | \cdot | \hat{\mathbf{z}}_t |}
\label{eq:cosine_sim}
\end{equation}
where $\mathbf{z}_t$ and $\hat{\mathbf{z}}_t$ are the ground-truth and predicted multiview representations at time step $t$. In addition to point-wise similarity, we also analyze statistical trends (mean, standard deviation, and maximum) of both original and predicted features to assess the model's ability to preserve global dynamics across time windows.

\subsection{Training analysis}
\label{traning_analysis}
\begin{figure}[h]
\centering
\includegraphics[scale=0.07]{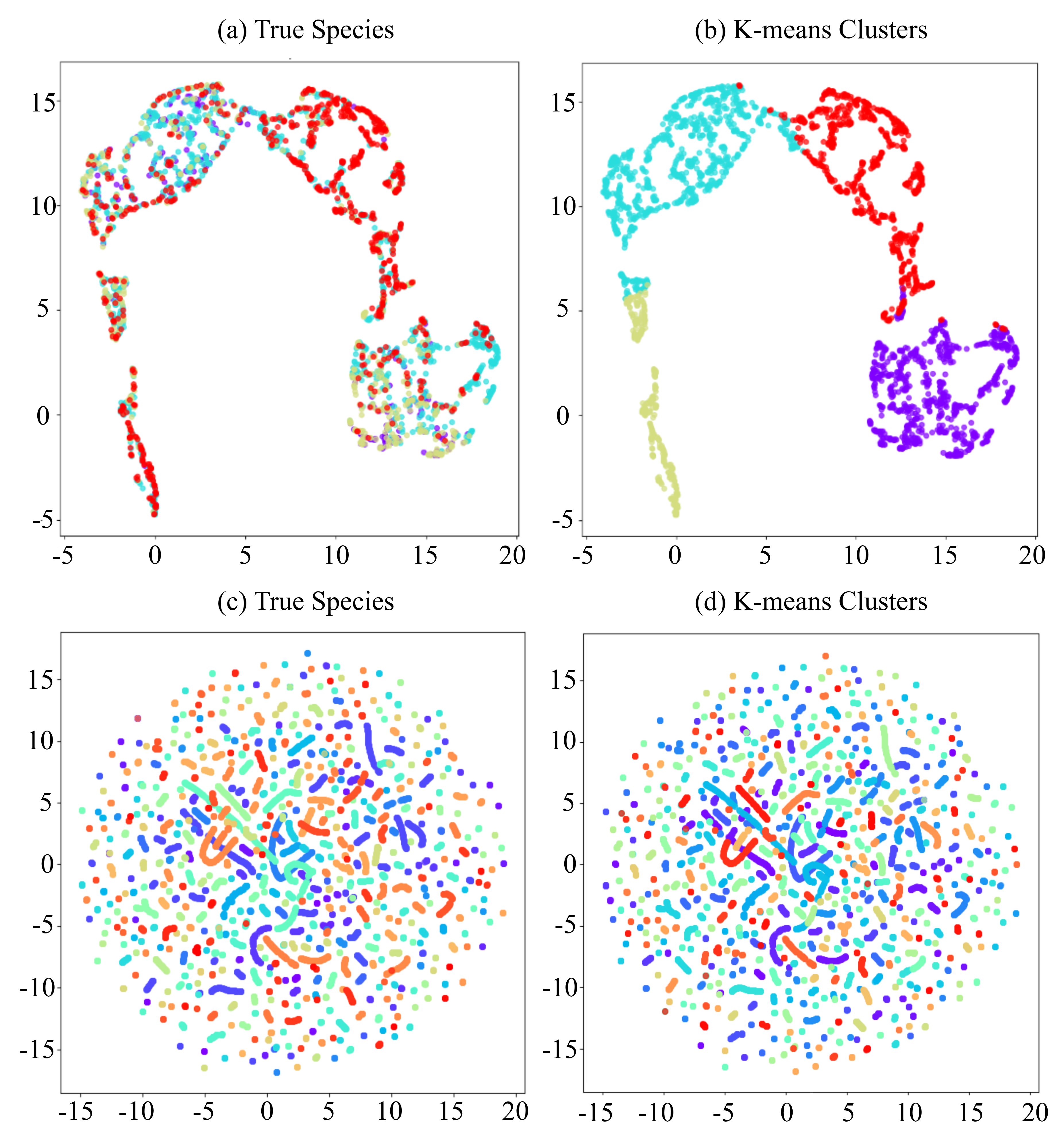}
\caption{Embeddings were clustered following contrastive learning training. Embeddings of five species (a) and their corresponding clusters (b) are shown for the XC-BS5 dataset, along with feature embeddings of 85 species (c) and their corresponding clusters (d) identified in the XC-British dataset.}
\label{fig:embeddings}
\end{figure}

The training process involved two stages: self-supervised representation learning using contrastive learning, followed by downstream tasks including species classification and future frame prediction. For contrastive learning, a Transformer-based encoder was trained on feature sequences to learn temporally-aware, discriminative representations of birdsong. Sinusoidal positional encoding, along with similarity embeddings, preserved sequence order (see Figure \ref{fig:embeddings}).
Training used the Normalized Temperature Scaled Cross Entropy Loss (NT-Xent) loss with a temperature $\tau$ of 0.07, optimized using Adam for 300 epochs with a learning rate of 1e-3, batch size of 64, and an exponential learning rate scheduler with gamma of 0.95 (see Section \ref{ablation_study}). 

The final contrastive losses were 0.3695 for the XC-British dataset and 0.3812 for the XC-BS5 dataset. As mentioned in Section \ref{dataset}, the subsets XC A-M and XC N-Z were excessively large. Due to resource constraints and class imbalance, we treated each alphabetical group as a separate subset and extracted features independently for each. After preprocessing, covering chromagrams, mel-frequency cepstral coefficients, and spectral descriptors, the features of each alphabetical group were merged into training sets. We achieved contrastive losses of 0.4261 and 0.3989 for the XC subsets A-M and XC N-Z, respectively. 

\begin{figure}[h]
\centering
\includegraphics[scale=0.28]{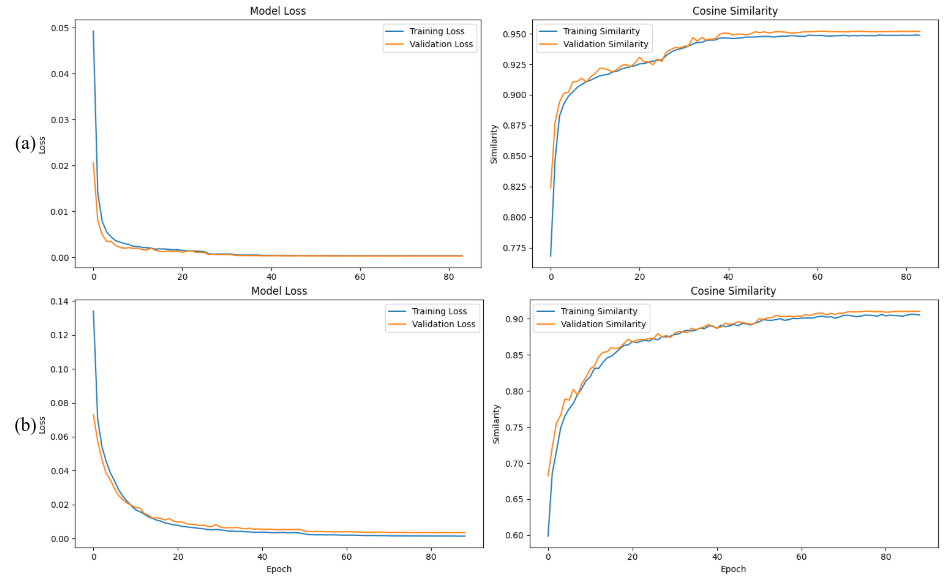} 
\caption{Training loss and cosine similarity trends for the temporal Transformer model evaluated on the (a) XC-British and (b) XC-BS5 datasets. For the XC-British dataset, the model achieved a validation cosine similarity of 0.9520 and an Mean Absolute Error of 0.0097. On the XC-BS5 dataset, the validation cosine similarity was 0.9103 with an Mean Absolute Error of 0.0285.}
\label{fig:training_cruve}
\vspace{-20pt}
\end{figure}

In the downstream stage, frozen encoder embeddings were used in two tasks. For species classification, we have selected a Random Forest with 100 estimators among four other state-of-the-art machine learning classifiers based on empirical results (see Section \ref{ablation_study}). For temporal modeling, a smaller Transformer was introduced to predict the next chromagram frame using mean squared error loss over 300 epochs, with a batch size of 32 and a learning rate of 1e-4. Figure \ref{fig:training_cruve} shows the model losses and cosine similarities for the XC-British and XC-BS5 datasets. Since early stopping mechanisms were incorporated to mitigate potential overfitting, optimal results were typically achieved in fewer than 100 epochs in the temporal model for future frame prediction. Table \ref{tab:training_params} details the training parameters. 

\begin{table}[!ht]
\centering
\caption{Training configuration for self-supervised contrastive pretraining and downstream tasks. Transformer settings are noted as: (blocks B$\times$attention heads H$\times$dimension D). The Normalized Temperature Scaled Cross Entropy Loss (NT-Xent) is associated only in the pretraining stage.}
\vspace{-5pt}
\label{tab:training_params}
\small
\begin{tabular}{l|l|l}
\midrule
\textbf{Stage} & \textbf{Parameter} & \textbf{Value} \\
\midrule
\multirow{8}{*}{\rotatebox{90}{Pretraining}} 
& Encoder & Transformer (4B$\times$4H$\times$128D) \\
& Dimension & 128 \\
& Feed Forward Net & 512 \\
& Loss & NT-Xent ($\tau$ = 0.07) \\
& Optimizer & Adam \\
& Learning Rate & 1e-3 \\
& Epochs & 300 \\
& Batch & 64 \\
& Scheduler & Exponential decay ($\gamma$ = 0.95) \\
\midrule
\multirow{7}{*}{\rotatebox{90}{Downstream}}  
& Classifier & Random Forest  \\
& Temporal Model & Transformer (2B$\times$2H$\times$64D) \\
& Loss & Mean Squared Error \\
& Epochs & 300 \\
& Batch & 32 \\
& Learning Rate & 1e-4 \\
& Early Stopping & Optimal $<$ 100 epochs \\
\midrule
\end{tabular}
\end{table}

\subsection{Classification evaluation}
We evaluated the proposed self-supervised framework on four diverse birdsong datasets. XC-British, XC-BS5, XC A–M, and XC N–Z. Performance is assessed using both general classification metrics and fine-grained diagnostic measures to offer a comprehensive view of predictive capability and model reliability. On the other hand, the XC-British dataset achieves the highest performance across nearly all metrics, including accuracy (98.41\%), F1-score (97.84\%), and Cohen’s Kappa (98.39\%). It also shows the lowest contrastive loss of 0.3695, which indicates a strong representation of learning from the unlabeled audio. Conversely, XC A–M and XC N–Z exhibit slightly lower yet competitive performance with F1-scores of 91.29\% and 90.94\%, respectively. Table \ref{tab:key_scores} reports the general classification metrics.

\begin{table}[!ht]
\centering
\caption{Summary of key classification metrics, including accuracy, Mean Absolute Error (MAE), Cohen's Kappa, Matthews Correlation Coefficient (MCC), and contrastive loss. The contrastive loss refers to the normalized temperature-scaled cross-entropy, which quantifies alignment between augmented views in the self-supervised framework.}
\label{tab:key_scores}
\footnotesize
\renewcommand{\arraystretch}{1.3} 
\resizebox{\columnwidth}{!}{%
\begin{tabular}{lcccc}
\midrule
\textbf{Metric} & \textbf{XC-British} & \textbf{XC-BS5} & \textbf{XC A-M} & \textbf{XC N-Z} \\
\midrule
Accuracy            & 98.41\% & 93.07\% & 91.89\% & 91.58\% \\
MAE                 & 0.2528  & 0.2635  & 0.3028  & 0.3112  \\
Cohen's Kappa       & 98.39\% & 93.12\% & 94.22\% & 93.61\% \\
MCC                 & 98.40\% & 91.90\% & 94.76\% & 94.12\% \\
Contrastive Loss    & 0.3695  & 0.3812  & 0.4261  & 0.3989  \\
\midrule
\end{tabular}%
}
\vspace{-15pt}
\end{table}

\begin{table}[h]
\centering
\caption{Core classification metrics across datasets. All results are presented in \%.}
\vspace{-5pt}
\label{tab:core_metrics}
\small
\resizebox{\columnwidth}{!}{%
\begin{tabular}{lcccc}
\toprule
\textbf{Metric}     & \textbf{XC-British} & \textbf{XC-BS5} & \textbf{XC A-M} & \textbf{XC N-Z} \\
\midrule
Precision           & 97.56 & 95.00 & 90.75 & 90.22\\
Recall              & 98.35 & 93.29 & 91.84 & 91.76\\
F1-score            & 97.84 & 94.10 & 91.29 & 90.94\\
Specificity         & 99.98 & 93.88 & 92.84 & 92.51\\
\bottomrule
\end{tabular}
}
\end{table}

The XC-BS5 dataset, with moderate species diversity and controlled background conditions, achieves an F1-score of 94.10\%. Although its accuracy (93.07\%) and recall (93.29\%) are lower than the XC-British dataset, it achieves higher Precision (95\%), which suggests a lower FPR on average. In addition, the tight agreement between Cohen’s Kappa and Matthews Correlation Coefficient in all settings further suggests consistent model behavior beyond chance. As reported in Table \ref{tab:core_metrics}, the model also consistently achieves high precision, recall, F1-score, and specificity, with the XC-British dataset achieving near-perfect performance: 97.84\% F1-score and 99.98\% specificity. The F1-score deviation across datasets remains below 3\%.

We also report complementary reliability metrics such as NPV, FPR, FDR, and FNR in Table \ref{tab:error_metrics}. The NPV exceeds 93\% across all datasets and peaks at 99.98\% on XC-British. Notably, FPR remains below 3.2\% across datasets, confirming a low rate of incorrect positive predictions even under noisy acoustic conditions. FNR also remains under 10.5\%, with the XC-British dataset exhibiting the lowest error rates.
\begin{table}[h]
\centering
\small
\caption{Error-related metrics across datasets, where $\uparrow$ indicates that higher values are better and $\downarrow$ indicates that lower values are better.}
\vspace{-5pt}
\label{tab:error_metrics}
\resizebox{\columnwidth}{!}{%
\begin{tabular}{lcccc}
\toprule
\textbf{Dataset} & \textbf{NPV $\uparrow$} & \textbf{FPR $\downarrow$} & \textbf{FDR $\downarrow$} & \textbf{FNR $\downarrow$} \\
\midrule
XC-British & 99.98\% & 0.0002 & 0.0126 & 0.0165 \\
XC-BS5     & 94.54\% & 0.0150 & 0.0926 & 0.0671 \\
XC A-M    & 93.74\% & 0.0280 & 0.1360 & 0.1050 \\
XC N-Z    & 93.23\% & 0.0315 & 0.1401 & 0.0824 \\
\bottomrule
\end{tabular}
}
\vspace{-15pt}
\end{table}

\subsection{Future frame prediction task}
Beyond species classification, future frame prediction offers several practical benefits in ecological and acoustic monitoring. First, it can simulate missing data recovery in field recordings, where environmental factors or sensor failure often cause audio dropouts. Second, predictive modeling can support the denoising or enhancement of incomplete sequences by anticipating expected harmonic structures. Finally, forecasting future vocalizations may support behavioral modeling, such as detecting call sequences, diurnal activity patterns, or anomalous disruptions in species-specific rhythms. These applications demonstrate the broader utility of temporally-aware representation learning, particularly in real-world monitoring deployments.

\subsubsection{Frame prediction and evaluation}
We evaluated the model’s ability to predict future frames using cosine similarity and Mean Absolute Error as primary metrics. The results show a strong predictive performance, with cosine similarity scores above 88\% for all datasets. The highest performance was observed on the XC-British dataset, where the model achieved a cosine similarity of 0.9520 and an Mean Absolute Error of 0.0097. Performance slightly decreased for the XC-BS5, XC A–M, and XC N–Z datasets, with cosine similarities ranging from 88.89\% to 91.03\% and Mean Absolute Error values between 0.0285 and 0.0346. The results are summarized in Table \ref{tab:future_frame_basic}.
\begin{table}[!ht]
\centering
\small
\caption{Future-frame prediction performance across datasets using cosine similarity and mean absolute error (MAE).}
\label{tab:future_frame_basic}
\begin{tabular}{lcc}
\toprule
\textbf{Dataset} & \textbf{Cosine Similarity $\uparrow$} & \textbf{MAE $\downarrow$} \\
\midrule
XC-British & 0.9520 & 0.0097 \\
XC-BS5     & 0.9103 & 0.0285 \\
XC A–M     & 0.8921 & 0.0334 \\
XC N–Z     & 0.8889 & 0.0346 \\
\bottomrule
\end{tabular}
\vspace{-10pt}
\end{table}

Further analysis of the distributional statistics of the original and predicted frames is presented in Table \ref{tab:future_frame_distribution}. Across the datasets, the predicted mean values closely align with the original means, with percentage differences below 1.6\%. Similarly, the standard deviation differences remain below 3.5\%. The maximum values and their deviations also remain tightly matched, with percentage differences below 1.5\%. The consistently low percentage differences between the mean and maximum statistics affirm the model's ability to generalize well to different datasets, despite some natural variability in species and recording conditions.

\begin{table*}[h]
\centering
\caption{Comparison of original and predicted distribution statistics for future-frame prediction, showing mean and max groups with absolute percentage differences.}
\label{tab:future_frame_distribution}
\footnotesize
\renewcommand{\arraystretch}{1.3}
\begin{tabular}{lccccccc}
\toprule
\multirow{2}{*}{\textbf{Dataset}} & 
\multicolumn{4}{c}{\textbf{Mean}} & 
\multicolumn{3}{c}{\textbf{Max}} \\
\cmidrule(lr){2-5} \cmidrule(lr){6-8}
 & Orig Mean ± SD & Pred Mean ± SD & Mean $\Delta$ (\%) & Std $\Delta$ (\%) & Orig Max ± SD & Pred Max ± SD & Max $\Delta$ (\%) \\
\midrule
XC-British & 0.3040 ± 0.1605 & 0.3076 ± 0.1629 & 1.18\% & 1.50\% & 0.5972 ± 0.3888 & 0.6060 ± 0.4004 & 1.47\% \\
XC-BS5     & 0.2911 ± 0.1603 & 0.2927 ± 0.1567 & 0.55\% & 2.25\% & 0.5815 ± 0.4065 & 0.5782 ± 0.4192 & 0.57\% \\
XC A–M     & 0.2848 ± 0.1532 & 0.2879 ± 0.1585 & 1.09\% & 3.46\% & 0.5612 ± 0.3976 & 0.5678 ± 0.4105 & 1.17\% \\
XC N–Z     & 0.2813 ± 0.1494 & 0.2856 ± 0.1520 & 1.53\% & 1.74\% & 0.5497 ± 0.3841 & 0.5531 ± 0.3973 & 0.62\% \\
\bottomrule
\end{tabular}
\end{table*}

\subsubsection{Case studies: future frame prediction across musical pitch classes}
\begin{figure*}[!ht]
\centering
\includegraphics[scale=0.037]{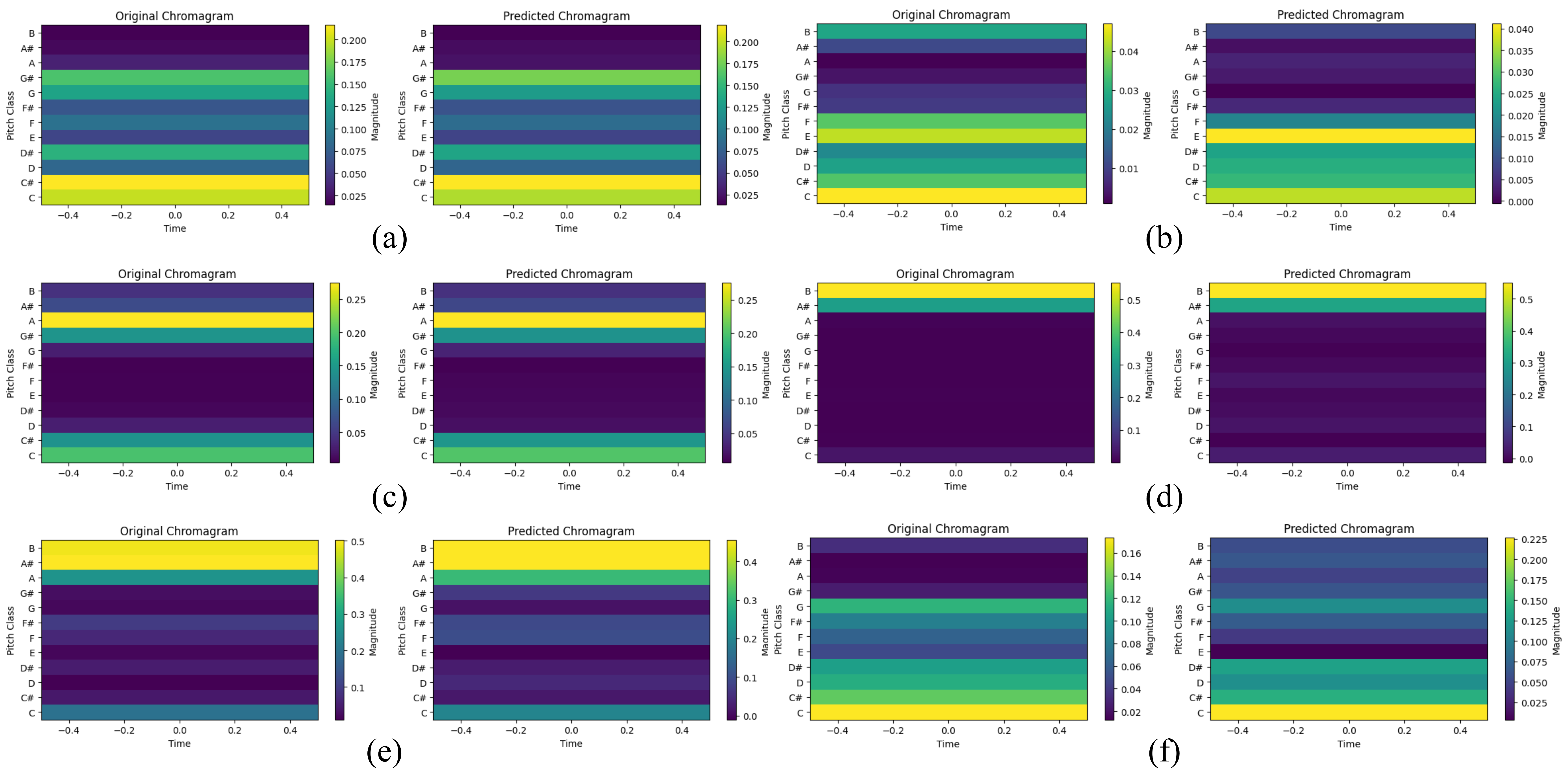} 
\caption{Predicted vs. original future-frame activations across 12 pitch classes for examples (a)–(f). Each subplot compares the spectral structure of the predicted frame (to the right) with the original frame (to the left). High-correlation examples (e.g., a, c, d) show strong alignment, while lower-correlation examples (b, f) exhibit modest divergence yet retain core spectral features.}
\label{fig:pitch_vis}
\vspace{-15pt}
\end{figure*}

We conducted a case study to assess how accurately our model predicts the future frame based on preceding audio, across 12 musical pitch classes (see footnote in Section \ref{preproc} for the classes). Six representative examples (a–f) were selected to compare predicted vs. original frame statistics and their correlations. 
As seen in Table \ref{tab:future_frame_case_study}, high-performing examples such as (a), (c), and (d) achieved correlations above 0.99, with minimal deviation in mean and max activation values, indicating strong temporal modeling. Example (e) also showed high fidelity (0.9851) despite a slight underestimation in peak energy. In contrast, examples (b) and (f) had lower correlations (0.8998 and 0.8598). Still, the predicted frames preserved the overall spectral structure. Figure \ref{fig:pitch_vis} visualizes the examples over the classes. 

\begin{table}[!ht]
\centering
\caption{Original vs. predicted future-frame statistics (examples a–f, musical pitch classes B–C).}
\label{tab:future_frame_case_study}
\small
\begin{tabular}{c c cc cc}
\toprule
\multirow{2}{*}{\textbf{Exp.}} & \multirow{2}{*}{\textbf{\textbf{Correlation}}} & \multicolumn{2}{c}{\textbf{Mean}} & \multicolumn{2}{c}{\textbf{Max}} \\ \cmidrule(lr){3-4} \cmidrule(lr){5-6}
           &                & \textbf{Orig} & \textbf{Pred}     & \textbf{Orig} & \textbf{Pred} \\
\midrule
(a) & 0.9947 & 0.1012 & 0.0997 & 0.2162 & 0.2187 \\
(b) & 0.8998 & 0.0225 & 0.0161 & 0.0470 & 0.0412 \\
(c) & 0.9987 & 0.0790 & 0.0807 & 0.2742 & 0.2751 \\
(d) & 0.9986 & 0.0764 & 0.0776 & 0.5503 & 0.5499 \\
(e) & 0.9851 & 0.1483 & 0.1467 & 0.5016 & 0.4549 \\
(f) & 0.8598 & 0.0766 & 0.0891 & 0.1734 & 0.2261 \\
\bottomrule
\end{tabular}
\end{table}
\subsection{Ablation studies}
\label{ablation_study}
All experiments in the ablation experiments were conducted over a reduced training budget of 50 epochs on the XC-British dataset, except for augmentation experiments, which were run for 100 epochs to better evaluate representation quality under contrastive objectives.

\begin{table}[h]
\centering
\caption{Ablation of domain-specific augmentations over 100 training epochs.}
\vspace{-5pt}
\label{tab:aug_ablation}
\resizebox{\columnwidth}{!}{%
\begin{tabular}{cccccc}
\toprule
\textbf{Pitch Shift} & \textbf{Time Mask} & \textbf{Chromagram Mask} & \textbf{CL} $\downarrow$ & \textbf{Cosine Sim} $\uparrow$ \\
\midrule
\cmark & \cmark & \xmark & 0.4376 & 0.9221 \\
\cmark & \xmark & \cmark & 0.4443 & 0.9185 \\
\xmark & \cmark & \cmark & 0.4490 & 0.9136 \\
\cmark & \xmark & \xmark & 0.4598 & 0.9053 \\
\xmark & \xmark & \xmark & 0.4817 & 0.8890 \\
\rowcolor{gray!20} \cmark & \cmark & \cmark & \textbf{0.4207} & \textbf{0.9370} \\
\bottomrule
\end{tabular}
}
\vspace{-10pt}
\end{table}

\begin{table*}[h]
\centering
\caption{Ablation study of key training hyperparameters on the XC-British dataset over 50 epochs. The best experiments in each aspect are bolded.}
\label{tab:ablation_ox}
\small
\begin{tabular}{l|c|lcc}
\midrule
\textbf{Aspect} & \textbf{Experiment} & \textbf{Description} & \textbf{Train Accuracy} & \textbf{Train Loss} \\
\midrule
\textbf{Baseline}
&  & Learning Rate = 0.01     &  &   \\
&  & Batch Size = 32       &  &   \\
& --  & Temperature $\tau = 0.1$  & 0.8395 &  0.6744 \\
&  & Projection $d=128$       &  &   \\
&  & Dropout = 0.1 &  &\\
&  & Multi-Layer Perceptron     &  &\\
\midrule
\textbf{Learning Rate}
& 0 & 0.0001 & 0.8982 & 0.4432 \\
& 1 & \textbf{0.001} & 0.9186 & 0.4207 \\
& 2 & 0.01 & 0.8901 & 0.4598 \\
& 3 & 0.1 & 0.8542 & 0.4821 \\
& 4 & 0.0005 & 0.9125 & 0.4259 \\
& 5 & 0.005 & 0.9011 & 0.4380 \\
\midrule
\textbf{Batch Size}
& 0 & 16 & 0.8960 & 0.4410 \\
& 1 & 32 & 0.9085 & 0.4290 \\
& 2 & 48 & 0.9120 & 0.4250 \\
& 3 & \textbf{64} & 0.9186 & 0.4207 \\
& 4 & 128 & 0.9140 & 0.4235 \\
\midrule
\textbf{Temperature $\tau$}
& 0 & $0.1$ & 0.9052 & 0.4401 \\
& 1 & $0.3$ & 0.9140 & 0.4279 \\
& 2 & \textbf{$0.5$} & 0.9186 & 0.4207 \\
& 3 & $0.7$ & 0.9101 & 0.4312 \\
\midrule
\textbf{Projection $d$} 
& 0 & $128$ & 0.9107 & 0.4296 \\
& 1 & \textbf{$256$} & 0.9186 & 0.4207 \\
& 2 & $512$ & 0.9132 & 0.4270 \\
\midrule
\textbf{Dropout}
& 0 & 0.1 & 0.9113 & 0.4315 \\
& 1 & \textbf{0.2} & 0.9186 & 0.4207 \\
& 2 & 0.4 & 0.9087 & 0.4392 \\
\midrule
\textbf{Classifier}
& 0 & Logistic Regression & 0.8890 & 0.4560 \\
& 1 & K-Nearest Neighbors & 0.9001 & 0.4425 \\
& 3 & Multi-Layer Perceptron & 0.8732 & 0.5153 \\
& 2 & \textbf{Random Forest} & 0.9186 & 0.4207 \\
\midrule
\end{tabular}
\end{table*}

\vspace{0.02\linewidth}\noindent\textbf{Effect of Augmentation.} As noticed in the influence of domain-specific augmentations (see Table \ref{tab:aug_ablation}), we find that removing any augmentation, such as pitch shifting, time masking, or chromagram masking, leads to increased contrastive loss and decreased cosine similarity, which indicates degraded feature alignment and temporal coherence. The best performance is achieved when all three augmentations are used together. 

\vspace{0.02\linewidth}\noindent\textbf{Hyperparameter and Classifier Ablation.} To fine-tune critical hyperparameters, including learning rate (LR), batch size (BS), temperature ($\tau$), projection dimension ($d$), and dropout rate, we conducted an ablation study. The best performance was consistently achieved with a learning rate of 0.001, batch size of 64, temperature of 0.5, projection size of 256, and dropout of 0.2. 

Additionally, we compared downstream classifiers and found that Random Forest outperformed logistic regression, K-Nearest Neighbors, and Multi-Layer Perceptron. Table \ref{tab:ablation_ox} summarizes the effect of hyperparameter selection based on empirical result analysis. 

\vspace{-15pt}
\subsection{Comparison with state-of-the-art models}
Unlike many prior approaches that primarily adopt supervised learning pipelines with limited feature sets, ARIONet integrates four complementary strategies: self-supervised learning using unlabeled data, temporal sequence modeling, future-frame prediction as an auxiliary task, and multi-feature fusion incorporating spectral, harmonic, and temporal descriptors.

Models such as \cite{li2025multi, xie2022multi, han2024bird} report high performance on curated datasets but do not consider unlabeled training or predictive learning objectives. Others \cite{wu2024orchard, wang2023hierarchical} incorporate hierarchical or multimodal cues but still rely on fully supervised data. A few works attempt semi-supervised learning, yet performance drops significantly when scaling to larger or noisier datasets.

\begin{table*}[h]
\centering
\small
\caption{Comparison of state-of-the-art birdsong classification methods. The final four columns indicate whether the model incorporates: (1) unlabeled/self-supervised training (\textbf{Unlabeled}), (2) temporal modeling (\textbf{Temporal}), (3) future frame prediction (\textbf{Future FP}), and (4) multi-feature fusion (\textbf{Multi-feature}). The results are presented in terms of accuracy (acc.), mean average precision (mAP), and F1-score. Our proposed model combines all four aspects and achieves competitive performance across both small-scale and large-scale bird datasets.}
\label{tab:birdsound_results}
\begin{tabular}{cc|lcccccl}
\toprule
\textbf{Ref.} & \textbf{Year} & \textbf{Dataset} & \textbf{Species} & \textbf{Unlabeled} & \textbf{Temporal} & \textbf{Future FP} & \textbf{Multi-feature} & \textbf{Result (\%)} \\
\midrule
\cite{liu2022birdsong}    & 2022 & Xeno-Canto & 4 & \xmark & \cmark & \xmark & \cmark &  \textit{mAP.} 95.90 \\
\cite{ugarte2024unveiling} & 2024 & Colombian Bird & 8 & \xmark & \cmark & \xmark & \cmark & \textit{acc.} 95.00\\
\cite{wu2024orchard} & 2024 & Orchard-birds & 10 & \cmark & \xmark & \xmark & \cmark & \textit{acc.} 99.40 \\
\cite{hu2023deep} & 2023 & UrbanSound8K & 10 & \xmark & \xmark & \xmark & \cmark & \textit{acc.} 98.34 \\
\cite{wang2023hierarchical} & 2023 & UrbanSound8K & 10 & \xmark & \xmark & \xmark & \cmark & \textit{acc.} 97.02 \\
\cite{duan2024sialex} & 2024 & UrbanSound8K & 10 & \xmark & \xmark & \xmark & \xmark &  \textit{acc.} 96.04 \\
\cite{hu2023deep} & 2023 & Huabei & 15 & \xmark & \xmark & \xmark & \cmark & \textit{acc.} 96.28 \\
\cite{xie2022multi} & 2022 & Xeno-Canto & 16 & \xmark & \xmark & \xmark & \cmark &  \textit{acc.} 96.25 \\
\cite{wu2024orchard} & 2024 & Birdsdata & 20 & \cmark & \xmark & \xmark & \cmark & \textit{acc.} 92.67 \\
\cite{hu2023deep} & 2023 & Birdsdata & 20 & \xmark & \xmark & \xmark & \cmark & \textit{acc.} 96.66 \\
\cite{li2025multi} & 2025 & Birdsdata & 20 & \xmark & \xmark & \xmark & \cmark & \textit{acc.} 97.81 \\
\cite{wang2023hierarchical} & 2023 & Birdsdata & 20 & \xmark & \xmark & \xmark & \cmark & \textit{acc.} 95.19 \\
\cite{wang2025dbs} & 2025 & Birdsdata & 20 & \cmark & \xmark & \xmark & \cmark & \textit{acc.} 97.09 \\
\cite{duan2024sialex} & 2024 & Birdsdata & 20 & \xmark & \xmark & \xmark & \xmark &  \textit{acc.} 93.66 \\
\cite{xiao2022amresnet} & 2022 & Birdsdata & 20 & \xmark & \xmark & \xmark & \cmark &  \textit{acc.} 92.60 \\
\cite{zhong2020multispecies} & 2020 & Collected & 24 & \cmark & \xmark & \xmark & \cmark & \textit{auc.} 99.50 \\
\cite{han2024bird} & 2024 & BirdVox-70k-unit03 & 25 & \xmark & \xmark & \xmark & \cmark & \textit{acc.} 98.72 \\
\cite{wang2025dbs} & 2025 & Custom & 30 & \cmark & \xmark & \xmark & \cmark & \textit{acc.} 97.54 \\
\cite{wei2024advanced} & 2024 & Collected & 31 & \cmark & \xmark & \xmark & \cmark & \textit{prec.} 85.60 \\
\cite{quinn2022soundscape} & 2022 & Collected & 54 & \xmark & \cmark & \xmark & \xmark &  \textit{prec.} 94.00 \\
\cite{gupta2021comparing} & 2021 & Cornell Bird Challenge & 100 & \xmark & \xmark & \xmark & \xmark & \textit{acc.} 90.00 \\
\cite{roemer2021automatic} & 2021 & Bat Sonotypes & 274 & \xmark & \xmark & \xmark & \cmark & \textit{auc.} 99.00 \\
\cite{kumar2024improving} & 2024 & BirdCLEF 2021 & 397 & \xmark & \cmark & \xmark & \cmark & \textit{f1.} 73.70 \\
\cite{ghani2025impact} & 2025 & Xeno-Canto & 585 & \xmark & \cmark & \xmark & \xmark & \textit{mAP.} 71.00 \\
\midrule
\multirow{4}{*}{\rotatebox{90}{ARIONet}} & \multirow{4}{*}{2025} & XC-BS5 & 5 & \multirow{4}{*}{\cmark} & \multirow{4}{*}{\cmark} & \multirow{4}{*}{\cmark} & \multirow{4}{*}{\cmark} & \textit{acc.} 93.07 \\
               &  & XC-British & 85 & & &  & & \textit{acc.} 98.41 \\
               &  & XC N-Z     & 106 & & &  & & \textit{acc.} 91.58 \\
               &  & XC A-M     & 153 & & &  & & \textit{acc.} 91.89 \\
\bottomrule
\end{tabular}
\end{table*}

In comparison, ARIONet achieves 98.41\% accuracy on the XC-British dataset, which includes 85 bird species with real-world recording variability. On more challenging subsets such as XC A–M and N–Z, involving over 250 species with diverse acoustic conditions, it maintains competitive performance (91.89\% and 91.58\%, respectively). 

These results are in line with or exceed those of prior models designed for controlled settings. Performance on the smaller XC-BS5 dataset (93.07\%) further reflects its generalizability in low-resource cases. Table \ref{tab:birdsound_results} provides a detailed comparison between ARIONet and a wide range of recent birdsong classification models evaluated across different datasets and species scales.

\section{Discussion} 
\label{discussion}
The ability to automatically and accurately classify birdsong across hundreds of species using self-supervised learning offers promising ecological benefits. Improved species-specific representations can significantly aid conservation efforts by enabling long-term biodiversity monitoring with minimal manual intervention. This is especially valuable in regions experiencing rapid habitat degradation or climate-induced migration, where real-time species tracking can inform conservation policy and prioritize protective measures \cite{ross2023passive}. Furthermore, forecasting vocalization patterns through future frame prediction may offer insights into behavioral changes, such as altered circadian rhythms or seasonal calling behavior, which can serve as early indicators of environmental stressors. 

However, as with any artificial intelligence-driven surveillance or monitoring system, ethical considerations must be addressed. Misclassification of rare or endangered species in protected areas could lead to incorrect ecological conclusions or conservation actions. In addition, passive acoustic monitoring in shared environments may inadvertently record human voices or activity, raising concerns about privacy and surveillance \cite{silva2024ai}. These concerns emphasize the need for transparent model auditing, careful deployment policies, and collaboration with local ecological stakeholders to ensure responsible and beneficial use.

In this study, we introduce ARIONet, a self-supervised framework designed to capture both the acoustic identity and temporal dynamics of birdsong in a unified manner. Rather than relying on static features or extensive manual labeling, our approach models birdsong as a harmonic sequence that evolves over time. We introduce a dual learning strategy: contrastive learning to capture species-specific patterns and future frame prediction to understand how these patterns evolve. This allows the model to learn rich, temporally aware embeddings that are both discriminative and biologically meaningful.

The core contribution of our framework lies in its self-supervised learning architecture that integrates contrastive representation learning with future-frame temporal prediction. The contrastive component enables the model to learn species-specific, view-invariant embeddings by comparing augmented chromagram views. Moreover, the temporal prediction module trains the model to anticipate future chromagram states, thus encouraging the encoder to internalize pitch sequences and temporal structures. This combination ensures that the learned representations are robust and temporally expressive. Furthermore, the application of domain-specific multiview augmentations, including chromagram masking, pitch shifting, and time masking, allows the model to generalize across a wide spectrum of species and vocal conditions, without losing discriminative power.

Empirical evaluation in four diverse datasets, including XC-British, XC-BS5, and the two extended Xeno-Canto subsets, demonstrates that ARIONet consistently achieves state-of-the-art results. The framework delivers 98.41\% classification accuracy and Cohen's kappa of 98.39\% on the XC-British dataset; it maintains high cosine similarity (up to 95.20\%) and overly low mean absolute errors. Ablation experiments further confirm the necessity of dual objectives: removing either the contrastive or predictive component leads to noticeable performance degradation. Similarly, excluding any type of augmentation significantly reduces alignment quality and predictive fidelity. These findings underscore the synergistic impact of the design choices made in the model architecture. By integrating biological relevance with technical robustness, ARIONet holds strong potential for scalable, responsible biodiversity monitoring across diverse ecosystems.

Although our proposed model shows strong performance and efficiency across our study's scope, there is one specific limitation: our preprocessing steps discarded low-energy segments (below 5\% of the maximum energy) to avoid overfitting. Although this choice did not affect our results, the bio-acoustic signals in other domains should utilize adaptive filtering.

\section{Conclusion}
\label{conclusion}
We proposed ARIONet, a novel self-supervised framework that unifies contrastive learning and future-frame prediction to capture both species-specific acoustic signatures and their temporal evolution in birdsong. ARIONet learns directly from raw audio through biologically inspired augmentations and a transformer-based encoder. Our key contribution lies in jointly optimizing two complementary objectives: distinguishing between species via contrastive learning on augmented views and modeling the temporal advancement of bird vocalizations through future frame prediction. Extensive experiments on four diverse datasets validate the effectiveness of our framework. We achieved classification accuracies of 98.41\%, 93.07\%, 91.89\%, and 91.58\% on the XC-British, XC-BS5, XC A-M, and XC N-Z, respectively. In addition to species classification, the model's ability to predict future frames supports applications such as signal reconstruction and behavioral forecasting in ecological monitoring systems. In the future frame prediction task, the model reached cosine similarity scores of up to 95.2\% and maintained low mean absolute errors. Through its dual-objective formulation, multiview augmentation strategy, and consistent empirical performance, the proposed framework shows strong potential as a self-supervised approach for birdsong classification and future frame generation. Our future work will explore further ecological modeling use cases and ensure responsible deployment in real-world sensitive or shared environments.

\section*{Declarations}
\noindent
\textbf{Conflict of Interests.} The authors declare that they have no financial conflicts of interest that could have influenced this work.\\
\textbf{Ethics Approval and Consent to Participate.} No additional ethics approval or consent was required, as all samples are publicly available and properly licensed.\\
\textbf{Dataset Availability.} All datasets used in this study are publicly available. The British Birdsong Dataset (XC-British\footnote{\url{https://www.kaggle.com/datasets/rtatman/british-birdsong-dataset}}), the Bird Song Data Set (XC-BS5\footnote{\url{https://www.kaggle.com/datasets/vinayshanbhag/bird-song-data-set}}), the Xeno-Canto Bird Recordings Extended (A–M) (XC A-M\footnote{\url{https://www.kaggle.com/datasets/rohanrao/xeno-canto-bird-recordings-extended-a-m}}), and the Xeno-Canto Bird Recordings Extended (N–Z) (XC N-Z\footnote{\url{https://www.kaggle.com/datasets/rohanrao/xeno-canto-bird-recordings-extended-n-z}}) were all obtained from open-access repositories and can be access from the links in the footnote.

\end{document}